\documentclass[10pt,conference]{IEEEtran}

\usepackage[utf8]{inputenc}
\usepackage[T1]{fontenc}
\usepackage{cite}
\usepackage{amsmath,amssymb,amsfonts,amsthm}
\usepackage{graphicx}
\usepackage{xcolor}
\usepackage{braket}
\usepackage{hyperref}
\usepackage{subcaption}
\usepackage{float}
\usepackage{pgfplots}
\usepackage{xfrac}
\usepackage{tikz}
\usepackage{quantikz}
\usetikzlibrary{positioning,calc,arrows.meta}
\usepgfplotslibrary{groupplots}

\definecolor{uninablue}{RGB}{0, 51, 101}
\definecolor{uninamagenta}{RGB}{158, 57, 106}

\usepackage{listings}
\begin{document}

\title{
    Q2NS Demo: A Quantum Network Simulator \\Based on ns-3
}

\author{
\IEEEauthorblockN{
    Francesco Mazza,
    Adam Pearson,
    Marcello Caleffi,
    Angela Sara Cacciapuoti}

\IEEEauthorblockA{
    \textit{Quantum Internet Research Group, University of Naples Federico II, Italy}}
}

\maketitle

\begin{abstract}
Q2NS is an open-source quantum network simulator built on ns-3, the de facto standard for classical network simulation. By inheriting ns-3’s mature classical stack and event-driven execution model, \textit{Q2NS} enables faithful co-simulation of quantum-network dynamics and classical signaling, a core requirement for the functioning of any quantum network.
Its modular architecture is designed for extensibility, with pluggable quantum-state backends (state-vector, density matrix, stabilizer) and a clean separation between network control and node-level operations. Q2NS comes with a quantum network visualizer \emph{Q2NSViz}, supporting interactive inspection of both physical- and entanglement-induced connectivity graphs, helping users interpret protocol behavior and entanglement manipulation processes. 
We present a demonstration of Q2NS, highlighting its ability to capture and simulate the coexistence of quantum and classical communication. The proposed demonstration presents quantum communication scenarios of increasing complexity: from entanglement distribution basics to multipartite graph-state manipulation, complemented by pre-loaded examples in \textit{Q2NSViz} that require no prior quantum communication or coding experience. 
\end{abstract}

\begingroup
\renewcommand\thefootnote{}
\footnotetext{This work has been funded by the European Union under Horizon Europe ERC-CoG grant QNattyNet, n.101169850. Views and opinions expressed are however those of the author(s) only and do not necessarily reflect those of the European Union or the European Research Council Executive Agency. Neither the European Union nor the granting authority can be held responsible for them.}
\endgroup

\section{Introduction}
The Quantum Internet fundamentally redefines the very notion of networking \cite{CalCac-26,CacCal-26, CacCalIll-25}. It requires the effective orchestration of quantum entanglement, while tightly coordinating with the classical network stack. Consequently, the hardware and protocols underpinning the future Quantum Internet are expected to differ substantially from those of the classical Internet. Simulation is a vital tool for keeping the time and cost of developing this complex system feasible. Yet developing such a simulator is itself a challenging task.
Today's simulators tend to focus on realistic quantum simulation, abstracting the classical stack such that in-depth quantum-classical co-simulation is challenging. Some are specialized for specific domains, such as QKD, while others are closed-source, rendering adaptation to the community's diverse and evolving needs difficult.

This demonstration showcases Q2NS~\cite{q2ns-repo,q2ns-journal,q2ns-qcnc}, an open-source, general-purpose simulator developed with the express purpose of enabling realistic quantum-classical co-simulation, along with its no-coding required visualizer, Q2NSViz. It does this by extending ns-3~\cite{ns3}, a well-established classical network simulator, into the quantum domain allowing a variety of quantum state representations and quantum networking capabilities to integrate naturally with the classical stack. Q2NS comes with ready-to-use pedagogic examples out of the box, paired with a quantum network visualizer. 
With this demonstration, both code-free examples and increasingly more complex quantum networking scenarios are covered.

\begin{figure}[t]
    \centering
    \input{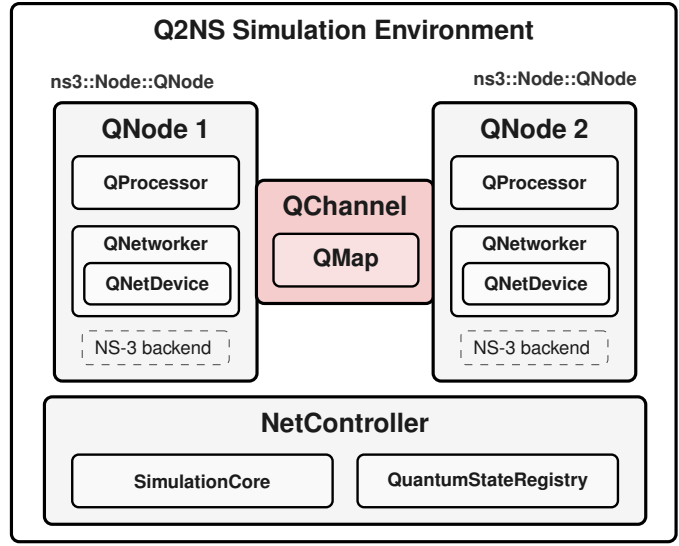}
    \caption{A high-level representation of Q2NS simulation environment, and its main hierarchies and entities. Figure reproduced from~\cite{q2ns-journal}.}
    \label{fig:architecture}
\end{figure}

\section{Q2NS in a Nutshell}
This section briefly summarizes Q2NS's design principles, modules, and capabilities, providing the minimum background needed to utilize and understand the simulation tool. 

\subsection{Architecture and philosophy}
Q2NS has a modular, extensible architecture designed to seamlessly integrate quantum and classical network paradigms. Its design is inspired by recent theoretical work on quantum network models and architectures \cite{CalCac-26,CacCal-26}. Due to its theory-grounded architecture and design, the simulator guarantees a rigorous modeling of entanglement-based quantum networks.

At the core of Q2NS architecture is a clear separation between \emph{network control} and \emph{in-network operations}, enabling flexible protocol composition while keeping well-defined roles and responsibilities among the components~\cite{q2ns-journal}. 

Q2NS extends ns-3 core network entities with specialized quantum counterparts: \textbf{\texttt{QNode}} inherits from \texttt{ns3::Node} -- equipped with a \textbf{\texttt{QNetDevice}} that inherits from \texttt{ns3::NetDevice} -- and \textbf{\texttt{QChannel}} inherits from \texttt{ns3::Channel}. 
Direct inheritance guarantees that each \texttt{QNode} is automatically able to engage in full classical protocol stack simulations -- including TCP/IP, sockets, routing, congestion control and much more -- without additional wiring.

While \texttt{QNode}s offer the interface for in-network quantum operations, the \textbf{\texttt{NetController}} is the centralized \textit{super-entity} ruling global state dependencies, which cannot be owned by single network objects~\cite{q2ns-journal}. This design enforces the separation between the \textit{operations} and the \textit{control}.

Fig.~\ref{fig:architecture} represents a simplified view of the simulation environment, highlighting the main entities and components in a two-node quantum network. An in-depth description of Q2NS' architecture is provided in~\cite{q2ns-journal}.

\subsection{Pluggable Quantum-State Backends}

Simulating quantum mechanical systems poses significant bounds, constraints, and limitations on both network scale and modeling accuracy. One quantum state representation may be particularly useful for a subset of networking scenarios, but ineffective for others. 
For this reason, Q2NS supports multiple quantum state representations through a unified plug-in interface. The users can select the most suitable \textit{backend} for their simulation, depending on their needs:

\begin{itemize}
    \item \textit{Ket} (statevector): for simulation of pure states;
    \item \textit{DM} (density matrix): for simulation of mixed states;
    \item \textit{Stab} (stabilizer): scalable Clifford-circuit simulation.
\end{itemize}

The same simulation code runs unchanged across all three backends. The backend can be set once on the \texttt{NetController}, without altering protocol logic~\cite{q2ns-journal}.

\subsection{Discrete Event Simulation Workflow}

Q2NS leverages the discrete-event simulation framework of ns-3, where all the events are scheduled and then executed in simulation-time order through a global event queue. This means that the calls are non-blocking and take the same form: a lambda wrapped in \texttt{Simulator::Schedule}. 

\begin{figure}[t]
    \centering
    \resizebox{\linewidth}{!}{%
    \begin{tikzpicture}[x=1cm,y=1cm,>=latex, font=\scriptsize]
            \node[draw,rectangle,text width=2.2cm,minimum height=1.5cm,align=center] (E1) at (0,0) {\textbf{Event 1}\\Prepare Bell pair\\t = 0 ms};
            \node[draw,rectangle,text width=2.2cm,minimum height=1.5cm,align=center] (E2) at (3,0) {\textbf{Event 2}\\Send half of pair\\ t = 1 ms};
            \node[draw,rectangle,text width=2.2cm,minimum height=1.5cm,align=center] (E3) at (6,0) {\textbf{Event 3}\\Prepare $|+\rangle$\\ t = 1 ms};
            \node[draw,rectangle,text width=2.2cm,minimum height=1.5cm,align=center] (EN) at (9,0) {\textbf{Event N}\\Apply corrections\\ t = 5 ms};

            \draw[->, thick] (E1) -- (E2);
            \draw[->, thick] (E2) -- (E3);
            \draw[->, thick, dotted] (E3) -- (EN);
            
        \end{tikzpicture}%
    }
    \caption{Event queue for quantum teleportation of a $\ket{+}$ state in Q2NS. Each protocol step is a timestamped callback.}
    \label{fig:teleportation}
\end{figure}
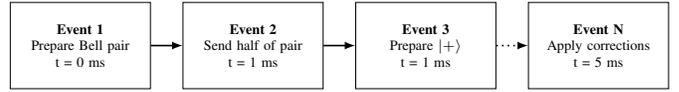

For example, the first steps of a teleportation protocol -- preparing a Bell pair, sending one half, and scheduling the Bell-state measurement -- can be expressed as a sequence of scheduled events, each firing after the previous one completes, as shown in Fig.~\ref{fig:teleportation}.
Each arrow in the figure corresponds to a \texttt{Simulator::Schedule} call that fires after its predecessor has completed\footnote{Qubit transmission is exposed directly through \texttt{A->Send(q,~B->GetId())}, and reception fires a callback registered with \texttt{B->SetRecvCallback}~\cite{q2ns-journal}.}.

\begin{figure*}
    \centering
    \includegraphics[width=0.95\linewidth]{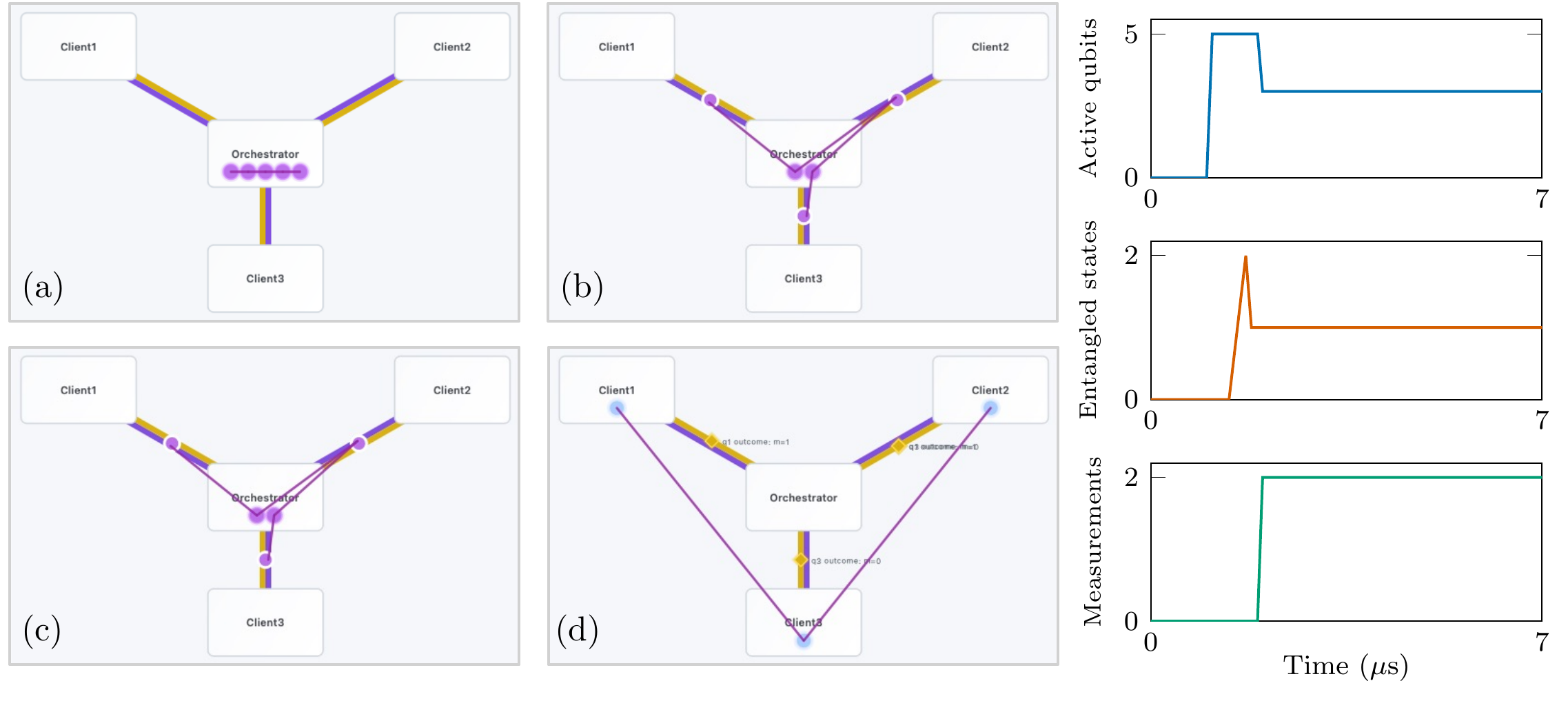}
    \caption{Q2NSViz rendering of the distributed graph-state protocol, including tracking of involved qubits and quantum states. The four snapshots show the protocol progression: initial qubit preparation, distribution along quantum channels (purple) and final measurements with classical correction outcomes delivered to the Clients.}
    \label{fig:demo}
\end{figure*}

\section{Demonstration}
\label{sec:demo}

The demonstration consists in two complementary parts. In the first one, pre-loaded simulation traces are explored in Q2NSViz, without coding required. In the second, progressively richer quantum network simulations are explored, revealing protocol logic and capabilities expressible with Q2NS. The demonstrated simulations range from basic Bell pair distribution to a full hybrid classical-quantum entanglement manipulation protocol, embedding quantum operations and UDP-based classical corrections.

\subsection{Protocol Visualization via Q2NSViz}\label{sec:demo-viz}
Q2NSViz is our visualization tool that replays a JSON trace supported by Q2NS during simulation runs. The interface presents quantum and classical channels, animated qubits as colored circles traveling along links, and entanglement graphs representing entangled quantum states. 
The traces supported by Q2NSViz annotate every gate or measurement (including entanglement generation and manipulation) and every classical and quantum information delivery. This allows users to follow the protocol events step by step or fast-forward to a specific moment in the timeline, turning the abstract event queue into a concrete, interactive visual narrative.

Four traces are pre-loaded and require no setup or simulation runs. 
The first two cover bipartite entanglement: (i) an all-to-all Bell pair distribution in a three-node network and (ii) a full quantum teleportation protocol run including Bell pair distribution, measurement, and classical UDP correction delivery over the classical channel. 
The remaining two examples involve multipartite entanglement states, namely, GHZ and cluster states. In the GHZ trace, a single source node prepares a 4-qubit state by applying a chain of two-qubit gates, then distributes the qubits to three remote parties waiting for acknowledgments to send the next qubit. The most complex trace shows a complete multipartite entanglement manipulation protocol, where a central node called \textit{Orchestrator} prepares and manipulates a 5-qubit cluster state. The orchestrator distributes part of the state to three remote nodes and measures its own qubits in the X-basis, resulting in the manipulation of the entanglement graph of the network, as depicted in Fig.~\ref{fig:demo}. The protocol is complete only after the classical correction messages are delivered to the clients.

\subsection{Interactive Simulation and Code Examples}\label{sec:demo-sim}

Q2NS is also equipped with a set of ready-to-use pedagogical examples.
Every Q2NS program starts with a \texttt{NetController}, which owns the quantum state registry and exposes the required methods to instantiate nodes and links. 
Setting up a two-node quantum network takes only a few lines:
\begin{lstlisting}
NetController net;
net.SetQStateBackend(QStateBackend::Stab);
auto A = net.CreateNode();
auto B = net.CreateNode();
auto ch = net.InstallQuantumLink(A, B);
ch->SetAttribute("Delay", TimeValue(NanoSeconds(10)));
\end{lstlisting}

From here, distributing a Bell pair $\ket{\Phi^\pm} = \sfrac{1}{\sqrt{2}}(\ket{00} \pm \ket{11})$ reduces to creating the entangled state locally and scheduling a send. \texttt{A->CreateBellPair()} prepares the state with a Hadamard and a CNOT, returning two qubit handles. The second is dispatched to \texttt{B} with \texttt{A->Send}, which is non-blocking: the qubit arrives only after the configured channel delay, upon which the registered callback fires. All protocol logic follows the same pattern. These mechanisms, familiar to ns-3 users, represent the basics to write hybrid quantum-classical protocols.

Channel noise is introduced via \textit{QMaps}, pluggable objects attached to \texttt{QChannel}s and applied automatically on qubit arrival. \textit{Qubit loss}, \textit{depolarizing}, and \textit{dephasing} maps are provided out of the box and can be composed. Adding noise to the Bell-pair example above immediately reveals its effect on measurement correlations without modifying protocol code.

Building on these primitives, a teleportation example shows the native integration with the classical layer. After a Bell-state measurement (\texttt{A->MeasureBell(q0,~q1)}), the sender packs the two outcome bits into a UDP datagram and sends it over a point-to-point classical link configured with standard ns-3 helpers. The receiver then applies the appropriate quantum corrections, triggered by the arrival of the classical packet with a specific callback registered on the classical socket. 

The final example scales to a cluster state on a star topology, where the stabilizer backend makes the scalability trade-off concrete: switching \texttt{QStateBackend::Stab} via a single call on the \texttt{NetController} allows the network to grow to tens or hundred of nodes with no change to the protocol. 

\section{Conclusion}
Q2NS provides a unified, open-source platform for studying hybrid quantum-classical network protocols. Building on the mature ns-3 ecosystem lowers the barrier to entry for classical networking researchers while giving quantum networking practitioners a physically grounded and extensible tool. The two-part demonstration -- trace visualization followed by guided simulation -- shows that protocols ranging from basic entanglement distribution to multipartite graph-state manipulation can be explored visually and analyzed concisely in code. 

\bibliographystyle{ieeetr}
\bibliography{biblio}

\end{document}